\documentclass[11pt,letterpaper]{article}
\usepackage[utf8]{inputenc}
\usepackage{url}
\usepackage{amsmath} 
\usepackage{graphicx}
\usepackage{amsfonts}
\usepackage{epstopdf}
\usepackage{amsopn}
\date{}

\usepackage[parfill]{parskip}

\graphicspath{{figs/}}

\renewcommand{\d}[2]{\frac{d #1}{d #2}} 
\newcommand{\pd}[2]{\frac{\partial #1}{\partial #2}} 
 
\newcommand{\sigmahat}{\hat{\sigma}} 
\newcommand{\muhat}{\hat{\mu}} 
\newcommand{\muhatmax}{\hat{\mu}_{\textrm{max}}} 

\newcommand{\add}[1]{{#1}}
\newcommand{\del}[1]{} 

\title{Why are U.S. Parties So Polarized? \\A ``Satisficing'' Dynamical Model}

\author{Vicky Chuqiao Yang\thanks{Santa Fe Institute, Santa Fe, NM}
	\and Daniel M.~Abrams\thanks{Department of Engineering Sciences and Applied Mathematics, and Northwestern Institute on Complex Systems, Northwestern University, Evanston, IL}
	\and Georgia Kernell\thanks{Departments of Communication and Political Science, University of California, Los Angeles, Los Angeles, CA}
	\and Adilson E. Motter\thanks{Department of Physics and Astronomy, and Northwestern Institute on Complex Systems, Northwestern University, Evanston, IL}}

\begin{document}
	
	\maketitle

	\begin{abstract}
		Since the 1960s, Democrats and Republicans in U.S. Congress have taken increasingly polarized positions, while the public's policy positions have remained centrist and moderate. We explain this apparent contradiction by developing a dynamical model that predicts ideological positions of political parties. Our approach tackles the challenge of incorporating bounded rationality into mathematical models and integrates the empirical finding of \textit{satisficing} decision making---voters settle for candidates who are ``good enough" when deciding for whom to vote. We test the model using data from the U.S. Congress over the past 150 years, and find that our predictions are consistent with the two major political parties' historical trajectory. In particular, the model explains how polarization between the Democrats and Republicans since the 1960s could be a consequence of increasing ideological homogeneity within the parties.
		
	\end{abstract}

	\section{Introduction} 
\begin{figure}[thb]
		\centering 
		\includegraphics[width = 0.6\textwidth]{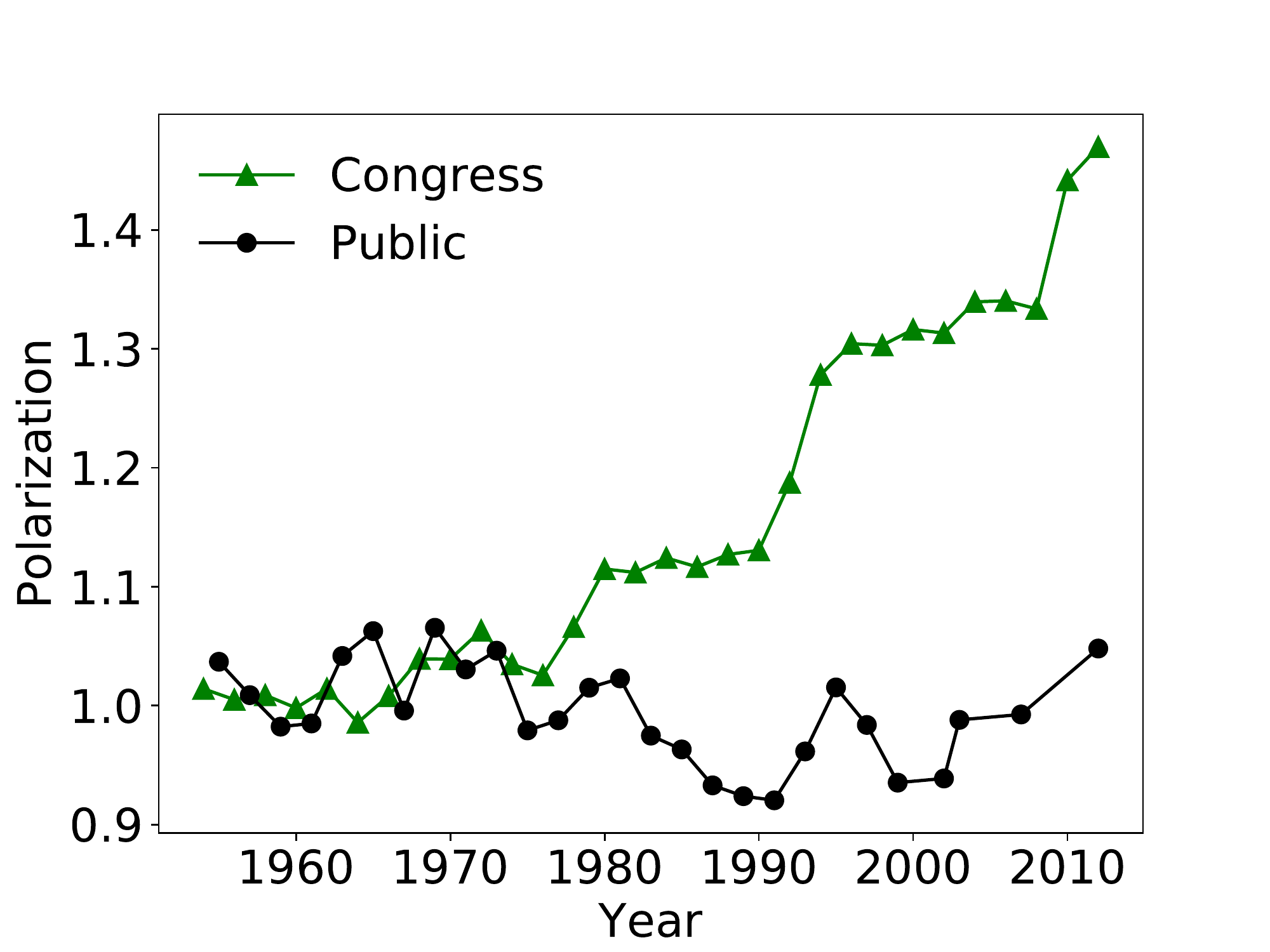}
		\caption{Polarization in the U.S. public and Congress (House and Senate combined), as measured by the standard deviation of ideological positions. The ideological positions are drawn from \cite{hill2015} for the public and \cite{dwnominateData} for Congress. Congress's standard deviation is scaled so that its mean for the first three data points matches that of the public. 
		}
		\label{fig:polar_contrast}
	\end{figure}

	The U.S. Democratic and Republican parties have polarized drastically since the 1960s \cite{andris2015, mccarty2016, poole1984}. Legislators' ideological positions are distributed bimodally, with an increasingly vast distance separating the two parties' modes \cite{baldassarri2008, hill2015}. Although what it means to be moderate has changed over the past half century, the general public's dispersion in ideology remains mostly the same: unimodal, centrist, and stationary \cite{fiorina2008_a, hetherington2009, hill2015}. Figure~\ref{fig:polar_contrast} illustrates contrasting ideological dispersion between the U.S. public and U.S. Congress. In a system designed to create democratic responsiveness, why have \add{members of congress} of the two major parties shifted while the general public appears to have remained unchanged? 
	
	Political party polarization is a complex social phenomenon. Recent research demonstrates that many key aspects of social processes can be explained by simple mathematical models. Examples include opinion dynamics \cite{castellano2009}, terrorist events \cite{clauset2007}, \add{the} spread of rumors \cite{borge2012}, and shifts in religious affiliation \cite{abrams2011}. A number of studies address aspects of voting and elections, such as consensus formation \cite{sood2005, xie2011}, \add{how opinion clusters form in the compromise process}\del{the compromise process} \cite{ben2003}, and how democratic voting can lead to totalitarianism \cite{galam1999}. Party polarization has now started to gain attention in the dynamical systems community. A recent study, for example, used a group competition model and found that the two major parties in the U.S. are increasingly benefiting from polarization in Congress \cite{lu2019}. 
	
	In political science, many theoretical models examine how individuals' voting behavior shapes political parties' positions. The classic Downsian model \cite{downs1957} assumes two-party competition in a one-dimensional ideology space, where citizens cast their vote for the ideologically closest party, and parties adjust their positions to maximize votes. The Downsian model predicts that both parties should converge to the median voter's position. In reality, however, the two major U.S. parties have dramatically polarized over the past half century.

	\add{One reason we believe the Downsian model fails empirical validation is the assumption that voters maximize their utility.} Empirical research suggests\del{, however,} that voters often behave \del{differently}\add{in ways that are only boundedly rational \cite{fisher2005}}. Instead of maximizing, people tend to {\sl satisfice}; that is, they accept what is ``good enough" and do not obsess over other options \cite{aumann1997,schwartz2002, simon1972, simon1985}. The decisions of a maximizer may approach those of a satisficer in the presence of noise, misinformation, missing information, decision fatigue, and other factors that limit one's ability to choose optimally. Moreover, experimental evidence demonstrates that for complex decisions, maximizers often make sub-optimal choices \cite{mao2016}. \add{Thus, even if some voters follow a maximizing decision-making strategy, in practice they will often fail to find the optimum and instead behave similarly to satisficers.}

	In this article, we (1) develop a simple mathematical model for political party positions, taking the more realistic satisficing decision making of voters into account; and (2) explain why political party polarization may develop in the absence of changes in voters' ideological positions. This work helps address the outstanding challenge in political polarization, and social systems more generally \cite{gintis2004}, of establishing unified mathematical models that are grounded in empirical findings. In addition to explaining key aspects of political polarization, our model's framework provides a foundation for future studies on other aspects of party dynamics, such as minor party influence and success.

	\section{Overview}
	Our modeling framework considers the voter population distributed in a one-dimensional ideology space, with two parties adjusting their positions along this continuum in order to win the most votes.  Voters decide which party to support based on satisficing---choosing one party randomly out of those that are satisfactory. The probability that a party is satisfactory to any given voter decays with the distance between that party and the voter, with the speed of decay determined by a parameter describing the party's inclusiveness, or tolerance of ideological diversity. We present a dynamical model for the parties' positions in ideology space. We then validate our model's predictions using empirical data on the distribution of U.S. legislators' ideological positions from 1861 to 2015, from which we estimate the position and inclusiveness of the Democratic and Republican party. The model employs as an input party inclusiveness, estimated from the data in each Congress, and predicts the corresponding position of each party over time. We then compare the predicted positions against historical positions of the two major U.S. parties, to establish a relationship between party polarization and party inclusiveness. Our model proposes a possible mechanism for the polarization of political parties without any change in the distribution of the public's ideological positions.
	

	\section{Deriving the mathematical model}
	
\begin{figure}[thb]
		\centering 
		\includegraphics[width = 0.7\textwidth]{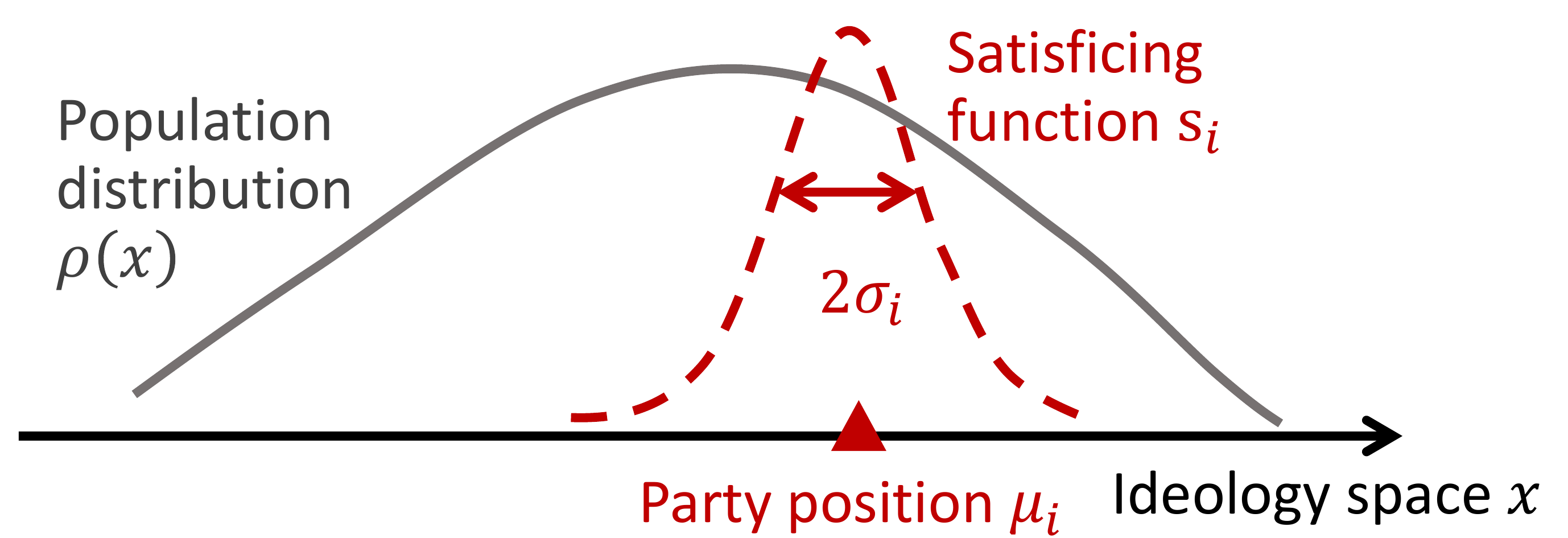}
		\caption{Sketch of the model setup showing the interpretation of parameters in the ideology space. The function $\rho$ (solid curve) represents the opinion distribution in the population of voters.  The parameter $\mu_i$ (triangle) marks the position of one out of possibly $n$ parties competing for voters, and function $s_i$ (dashed curve) represents that party's satisficing function. }
		\label{fig:intro}
	\end{figure}

	Using congressional roll call voting records, empirical research has found that the U.S. political division can be well represented by a single dimension \cite{poole1984}.\footnote{While a single dimension historically performs well at explaining the vast majority of legislative voting behavior, there are several notable time periods---such as during the civil rights movement---during which a second dimension is salient \cite{dwnominateData}.} We refer to this dimension as the \textit{ideology space}, with left being liberal and right being conservative. We also describe how liberal or conservative a party or voter is by their position in the same ideology space, henceforth referred to as \textit{ideological position}, or simply as \textit{position}. Survey data reveals that the American public's ideology is distributed in this space in a unimodal manner, peaked at a moderate position and well approximated by a Gaussian (see the supplementary material, section 1). Thus, we consider the voters to be distributed in the ideology space according to a Gaussian function $\rho (x)$. Without loss of generality, we set the mean of the Gaussian to $0$, and the standard deviation to $\sigma_0$. Each party $i$ adopts a position, denoted as $\mu_i$, along this ideology space. See Fig.~\ref{fig:intro} for a sketch of the model setup.
	
	
	Consider an election among $n$ parties. We assume the probability that a voter positioned at $x$ is satisfied with party $i$ decays with the voter's distance from $\mu_i$ and that this decay is symmetrical. We refer to this decaying probability as the \textit{satisficing function}, $s_i(d_i)$, where $d_i = |x-\mu_i|$. We also assume $s_i(0) = 1$ (meaning that a voter perfectly aligned with a party will be satisfied with that party), and $s_i(d_i) \to 0$ as $d_i \rightarrow \infty$. One function that satisfies these properties is, again, a Gaussian: $s_i(d_i) = \exp[- d_i^2/(2 \sigma_i^2)]$ (here scaled to have unit peak amplitude). The parameter $\sigma_i$ represents how tolerant voters are of parties with ideologies different from their own \add{(Fig.~\ref{fig:intro})}. It can also be interpreted as the \textit{inclusiveness} of party $i$; that is, how much a party appeals to voters at distant ideological positions. \add{Parties with lower inclusiveness are more ideologically homogeneous.} 
	
	We break down the satisficing decision-making process into explicit statements: 
	\begin{enumerate} 
		\item Voters who are satisfied with none of the $n$ parties abstain from voting. 
		\item Voters who are satisfied with a subset $M$ of the $n$ parties vote for each party in $M$ with probability $p_i$, where $\sum_M p_i = 1$. (Below, we assume for simplicity that $p_i$ is the same for all parties in this set, but this assumption is not essential.)
	\end{enumerate}
	
	We first examine the case in which two parties, party $1$ and party $2$, compete. Since the governing equations for the two parties are similar, in the following equations, we use index $i$ to denote either party 1 or 2, and $j$ the other party. The expected share of potential voters at position $x$ who vote for party $i$ can be expressed as
	
	\begin{equation}\label{eq:def_p}
	p_i(x| \mu_i, \mu_j)  = \underbrace {s_i(d_i) \left(1 - s_j(d_j) \right)}_\text{A} +   \underbrace {\frac{1}{2} s_i(d_i) s_j(d_j)}_\text{B} \;, 
	\end{equation}
	
	where term A represents the expected share of voters at position $x$ who are satisfied with party $i$ only, and term B represents half of those who are satisfied with both parties.
	The expected total number of voters for party $i$, which we denote $V_i$, is obtained by integrating $p_i(x| \mu_i, \mu_j)$ over the ideology space weighted by the population density $\rho$:
	\begin{equation}\label{eq:def_V}
	V_i(\mu_i, \mu_j) = \int_{-\infty}^{\infty} \rho(x) p_i(x|\mu_i, \mu_j) dx \;.
	\end{equation}

	To model dynamical changes in positions over time, we assume parties move in the direction that increases the number of votes received at a speed proportional to the potential gain:
	\begin{equation}\label{eq:dynam}
	\d {\mu_i} t  = k \pd{V_i}{\mu_i} \;,
	\end{equation}
	where $k$ is a positive constant determined from data that sets the time scale. Equation (\ref{eq:dynam}) generates predictions for party $i$'s position over time, where the left side represents the speed of position change and the right side represents the change in party's votes. While we derived this equation for the two-party case, it admits an immediate extension to the $n$-party case with the exact same form, except that the vote, $V_i$, is then defined relative to all $n$ parties. In supplementary material, section 4, we explore an alternative model where parties maximize vote \textit{share}. This model produces similar outcomes (although it requires additional complexity). 
	
	
	\section{Results and comparison with empirical data}
	\subsection{Fixed points of the system}
	\label{secSI:fixed_points}

	\begin{figure}
		\centering 
		\includegraphics[width = 1\textwidth]{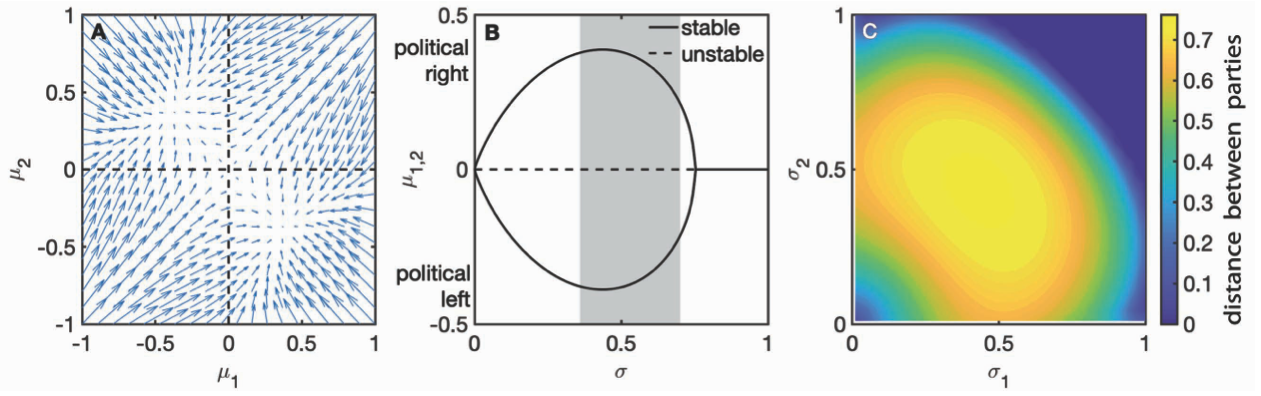}
		\vspace{-0.2in}
		\caption{(A) A vector field that demonstrates the movement of the two parties at various ideological positions. Arrows show the direction of the party positions' movement, with the length of each arrow proportional to the speed of movement. The parameters used in this figure are the inclusiveness parameters $\sigma_1 = \sigma_2 = 0.5$ and time scale constant $k = 1$. (B) Fixed points for party positions ($\mu$) as a function of the inclusiveness parameter of the two major parties, assuming symmetrical inclusiveness $\sigma_1 = \sigma_2 = \sigma$.  The stable fixed points are shown as solid curves, and the unstable fixed point is indicated by the dashed line. The shaded area indicates the inclusiveness parameter region corresponding to that \add{estimated from}\del{of} the DW-NOMINATE data discussed in section \ref{sec:data} using methods discussed in section \ref{sec:para}. (C) Polarization, as measured by the distance between parties at steady state, as a function of both $\sigma_1$ and $\sigma_2$. In all three panels, the parameter $\sigma_0$ is set to $0.93$, which is the same value used in Fig.~\ref{fig:result}.}
		\label{fig:modelBehavior}
	\end{figure}
	
	We solve for the fixed points of the system both analytically and numerically with analytical results presented in section \ref{subsec:ana_fp} and numerical results presented in Figs.~\ref{fig:modelBehavior} and \ref{fig:result}. The results reveal that for sufficiently large inclusiveness parameters, $\sigma_{1,2}$, the two parties are attracted to a single stable equilibrium: $\mu_1=\mu_2=0$, at the center of the ideology space. However, for smaller $\sigma_{1,2}$, the two parties stabilize at a finite separation. Fig.~\ref{fig:modelBehavior}A visualizes the dynamics of a two-party system by plotting the derivatives $d \mu_1/dt$ and $d \mu_2/dt$ in Eq.~\eqref{eq:dynam} as a vector field. In this example, the two parties stabilize at a finite separation. The fixed points and their stability are shown in Fig.~\ref{fig:modelBehavior}B for the symmetric case, $\sigma_1=\sigma_2$. We note that Fig.~\ref{fig:modelBehavior}B plots the positions of both parties on the same vertical axis. In the asymmetric case where $\sigma_1 \neq \sigma_2$, the polarization, as measured by the distance between the two parties at steady state, is a function of both $\sigma_1$ and $\sigma_2$. The numerical results for this case are shown in Fig.~\ref{fig:modelBehavior}C. The system equilibrates at a finite separation in most of the parameter space. 
	
	An intuitive understanding of this result is that when the inclusiveness parameters are large, few voters abstain. Thus, parties are attracted to the center of the ideology space through the same mechanism as in the classic Downsian model that predicts the convergence of both parties to the median voter \cite{downs1957}. However, as the inclusiveness parameters decrease, convergence to the center increases the number of voters at the tails of the ideology distribution abstaining from voting. In that case, the parties benefit from moving away from the center, with an equilibrium distance determined by the parameters. Although the model was presented in a one-dimensional ideology space, the model's behavior is similar in two dimensions (see supplementary material section1). 
	
	
		\subsection{Analytical results for two identical parties} \label{subsec:ana_fp}
		The right hand side of the system specified by Eq.~\eqref{eq:dynam} can be computed analytically for the symmetrical case $\sigma_1 = \sigma_2 = \sigma$: 
		\begin{equation} \label{eq:intRhs}
		\d{\mu_i}{t} = 
		\frac{ \sigma_0 [ \mu_i(\sigma^2 + \sigma_0^2) - \mu_j \sigma_0^2] }{2 \sigma(\sigma^2 + 2\sigma_0^2)^{3/2}} e^{-\frac{(\mu_i^2 + \mu_j^2)\sigma^2 + (\mu_i - \mu_j)^2 \sigma_0^2}{2 \sigma^2 (\sigma^2 + 2\sigma_0^2)}} 
		- \frac{ \mu_i \sigma_0 \sigma e^{- \frac{\mu_i^2}{2(\sigma^2 + \sigma_0^2)}} }{(\sigma^2 + \sigma_0^2)^{3/2}} 
		\;.
		\end{equation}
		
		Solving for the equilibria $d \mu_i/ dt = 0$, and imposing symmetry in party positions $\mu_2 = -\mu_1 = \mu$ (this follows from the symmetry in $\sigma_1 = \sigma_2$), we find
		\begin{equation}\label{eq:analyticFP}
		\mu \left[ -2 \sigma^2 (\sigma^2 + 2\sigma_0^2)^{1/2} e^{- \frac{\mu^2}{2(\sigma^2 + \sigma_0^2)}}  
		+ (\sigma^2 + \sigma_0^2)^{3/2}e^{-\frac{\mu^2}{\sigma^2}} \right]  = 0 \;.
		\end{equation}
		Clearly $\mu = 0$ is a solution, and thus $\mu^\star = 0$ is a fixed point of the system for all values of $\sigma$ and $\sigma_0$. Other possible fixed points are given by the solution of the equation
		$(\sigma^2 + \sigma_0^2)^{3/2}e^{-\frac{\mu^2}{\sigma^2}} - 2 \sigma^2 (\sigma^2 + 2\sigma_0^2)^{1/2} e^{- \frac{\mu^2}{2(\sigma^2 + \sigma_0^2)}} = 0$, which can be solved for $\mu^2$ explicitly as
		\begin{equation} \label{eq:mustar}
		\left(\mu^\star\right)^2 = \sigma^2 
		\left(\frac{\sigma^2 + \sigma_0^2}{\sigma^2 + 2 \sigma_0^2}\right)
		\ln \left[ \frac{\left(\sigma^2 + \sigma_0^2\right)^3} 
		{4 \sigma^4 \left(\sigma^2 + 2 \sigma_0^2 \right)}
		\right] \;;
		\end{equation}
		the solid line in Fig.~\ref{fig:modelBehavior}B is this curve.  Note that this expression can be rewritten in terms of nondimensional parameters $\{\hat{\sigma} \equiv \sigma / \sigma_0,~\hat{\mu} \equiv \mu/\sigma_0\}$ simply by setting $\sigma \mapsto \hat{\sigma}$ and $ \sigma_0 \mapsto 1$.
		
		The trivial solution $\mu^\star = 0$ is the only fixed point for $ \sigma > \sigma_c$ for some $\sigma_c$, and it is of interest to understand how $\sigma_c$ depends on system parameters\footnote{\add{The result that the trivial solution $\mu^\star = 0$ is the only fixed point for $ \sigma > \sigma_c$, holds even for two parties with $\mu_1 \neq \mu_2$, as can be seen with a series expansion for large $\sigma$ retaining only leading order terms.}}.  This critical value can be understood by noting that the nontrivial fixed points for $\mu$ given in Eq.~\eqref{eq:mustar} cease to exist when the logarithm becomes negative, i.e., when the argument of the logarithm becomes less than one; $\sigma_c$ is the critical value at which the argument is exactly one. Expressing this condition in the nondimensionalized form gives
		\begin{equation}
		\frac{\left(\sigmahat_c^2 + 1\right)^3} 
		{4 \sigmahat_c^4 \left(\sigmahat_c^2 + 2 \right)} = 1\;,
		\end{equation}
		and rearranging the terms leads to a cubic polynomial in $\sigmahat_c^2$:
		\begin{equation}
		3 \sigmahat_c^6 + 5 \sigmahat_c^4 - 3 \sigmahat_c^2 - 1 = 0\;.
		\end{equation}
		
		This equation has three real roots for $\hat \sigma^2_c$, at $\sigmahat_c^2 \approx \{-2.07, -0.247, 0.652\}$, and only the last root allows for a real-valued solution $\sigmahat_c \approx 0.807$.  The system undergoes a subcritical pitchfork bifurcation at this point.
		
		The exact shape of the fixed point curve is defined by Eq.~\eqref{eq:mustar}, which cannot be solved explicitly for $\sigma$ (or $\sigmahat$ in nondimensional form).  A convenient approximation can be written by expanding the right-hand side to the leading order in $\sigmahat$, yielding $ \muhat^2 \approx -\sigmahat^2 \ln (\sqrt{8} \sigmahat^2)$.  Define $W \equiv -\muhat^2/\sigmahat^2$, and rearrange the previous equation to get $-\sqrt{8} \muhat^2 \approx W \exp(W)$.  This equation is solved by $W = W(-\sqrt{8} \muhat^2)$, where $W$ is now identified as the special function known as the Lambert $W$ function. Thus $\sigmahat^2 \approx -\muhat^2 / W(-\sqrt{8} \muhat^2)$.
		
		The Lambert $W$ function is real-valued for negative arguments only when the argument has magnitude less than $1/e$. This sets a bound on the maximum $\muhat^2$ for which a solution exists since $\sqrt{8} \muhatmax^2 \approx 1/e $ implies $\muhatmax \approx 2^{-3/4} e^{-1/2} \approx 0.361$.  The corresponding value of $\sigmahat$ is then immediate since $W(-1/e)=-1$, yielding $\sigmahat_{\textrm{max}} \approx \muhatmax \approx 2^{-3/4} e^{-1/2}$.  These approximations are useful in understanding the general shape of the bifurcation curve, though the critical points of the exact expression in \del{function} Eq.~\eqref{eq:mustar} can be easily computed numerically.

	
	\subsection{Empirical Data} \label{sec:data}
	
	To examine changes in real-world party positions over time, we employ ideological positions for Democratic and Republican legislators calculated from their congressional roll call voting records using the Dynamic, Weighted, Nominal Three-Step Estimation (DW-NOMINATE) method \cite{dwnominateData}. DW-NOMINATE is a multi-dimensional scaling method that first calculates a pairwise distance for every two members of the U.S. House of Representatives based on similarities in their roll call voting records. It then projects the resulting high-dimensional network of representatives to a low dimension while preserving the pairwise distance relation as much as possible. The representatives' relative positions in this low-dimensional space are referred to as their ideology scores.\footnote{Note that this measure is relative in nature---it reflects the similarities and differences among members' voting records across a large number of bills, not their positions on any specific issue.} The same method can be used for scaling the positions of U.S. Senators, and in what follows we use a combined dataset of both representatives and senators (see supplementary material section 2 for data source). 
	
	As an example, Fig.~\ref{fig:result}A displays a histogram of Democratic and Republican legislators' ideological positions during the 2013-15 term, the most recent in the data. 
	
	\subsection{Model Validation}
	
\begin{figure}[!thb]
		\centering 
		\includegraphics[width = 0.9\textwidth]{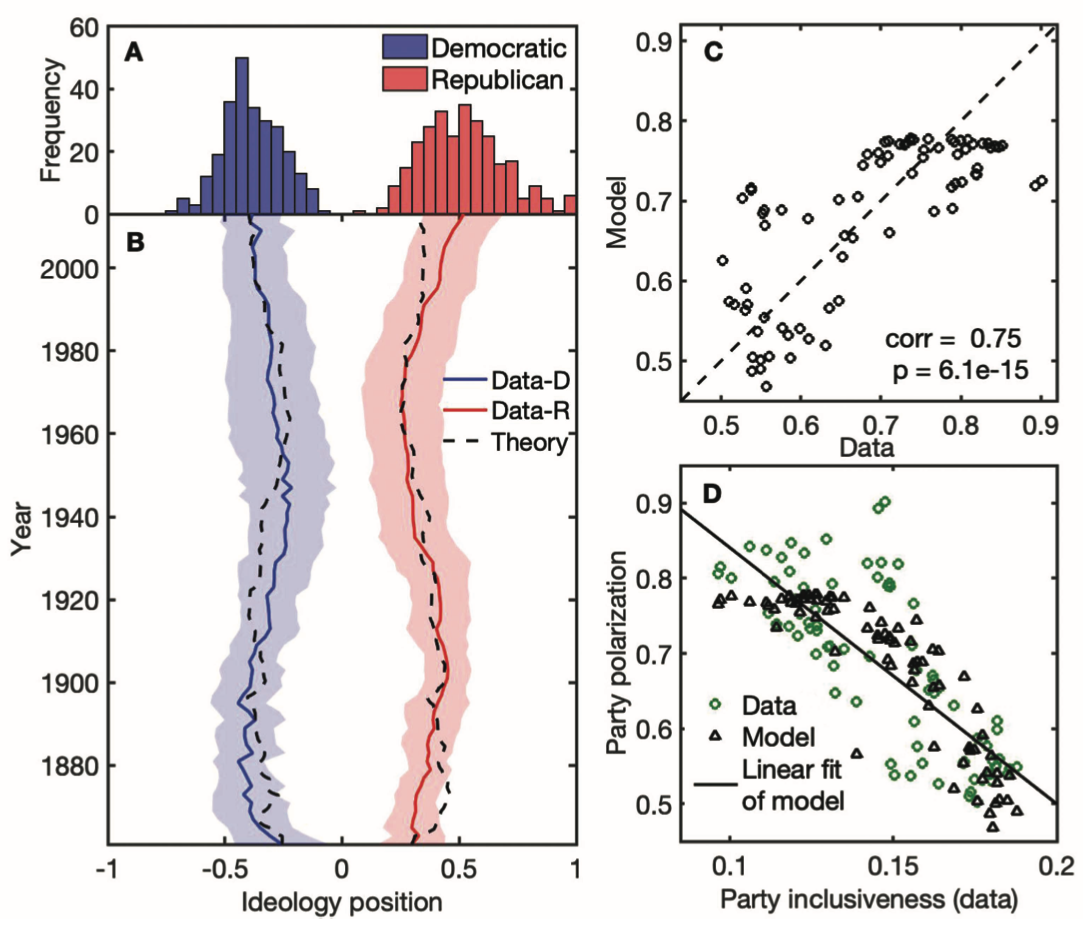}
		\caption{Predictions and validations of the satisficing model. (A) Histogram of ideological positions for the Democratic and Republican parties in the 2013-15 Congress. (B) Comparison of the model prediction with time series data for party positions. The solid curves indicate the empirical party positions, estimated by the mean position of the party's members of Congress, while the shaded areas mark the associated standard deviations. The dashed curves show our model predictions. (C) Comparison of party polarization predicted by our model and polarization observed in the data. The Pearson correlation between the prediction and the data is 0.75, with p-value $6.1\times 10^{-15}$. The dashed line corresponds to correlation 1 and is a reference. (D) Party polarization as a function of $\sigma$ (average of the Democratic and Republican parties' inclusiveness parameters). Each marker represents one Congress term. In both the model's prediction and the data, party inclusiveness is negatively associated with polarization. In panels (C) and (D), party polarization is measured as the distance between party positions.}
		\label{fig:result}
	\end{figure}

	For purposes of comparing our model with data, we take the mean of each party's distribution to represent its position, $\mu_i$, in the model and the standard deviation of each party's distribution to represent its inclusiveness parameter, $\sigma_i$. As members of Congress move farther from the party mean, we infer that the party is more inclusive---and may thus appeal more to voters farther away from its center. Repeating that procedure for all available Congresses leads to a series of party positions over time since 1861, shown as the solid curves in Fig.~\ref{fig:result}B. We note that two distinct polarization metrics---the distance between party means (used in Fig.~\ref{fig:result}) and the standard deviation of the legislators' ideology distribution (used in Fig.~\ref{fig:polar_contrast})---consistently demonstrate the striking increase in party polarization since the 1960s.
	
	Importantly, Fig.~\ref{fig:result}B also compares the model's predictions for the two major parties' positions to historical data. We set the two parties' positions in 1861 as their initial conditions and numerically simulate the dynamical system \eqref{eq:dynam} to predict their positions in subsequent years. In the simulations, we update the inclusiveness parameter according to the data every Congress, and the model outputs the positions of both parties according to Eq.~\eqref{eq:dynam}. Parameters $\sigma_0$, $k$, and a proportionality constant associating the standard deviation in the DW-NOMINATE score with the inclusiveness parameter are determined from the fit to the data. An animated visualization of Figs. 4A and 4B is included as a supplementary video. 
	
	Our theory shows good agreement with \add{the} data (Pearson correlation 0.75; see Figs.~\ref{fig:result}C and \ref{fig:result}D). Notable deviations are observed around the times of the First and Second World Wars and following the recent rightward move by the Republican Party, as shown in Fig.~4B. 
	
	A central prediction of our model is that lower inclusiveness (more homogeneity within parties) will lead to higher party polarization. We find support for this prediction in empirical data, as shown in Fig.~\ref{fig:result}D. This \add{finding} suggests that it is not necessary for the voting population to become polarized in order for increased political party polarization to occur. Indeed, the electorate need not change at all; it is held constant in our simulations. We also perform robustness checks using an alternative measure of party polarization as well as an independent data source for the inclusiveness parameter\del{,} and reach the same qualitative conclusions (see supplementary material section 3).

	\subsection{\add{Parameter fitting}} \label{sec:para}
	\add{When determining the parameters from the data,} we relate the unit\add{s} of the DW-NOMINATE ideology score\add{s} to units for variables in our model by assuming a linear scaling. For example, the party inclusiveness parameter, $\sigma_i$, is assumed to be linearly related to the DW-NOMINATE data through the scaling $\sigma_i = b \; \sigma_{\text{data}, i}$. We fitted the parameters in the model by minimizing the \add{1}\del{2}-norm for differences between the time series predicted by the model and the time series from data. Three parameters are fitted: (1) $\sigma_0$, the standard deviation of \add{the} population ideology distribution; (2) $b$, the proportionality constant relating the standard deviation in the DW-NOMINATE data with the party inclusiveness parameter in the model, $\sigma_i$; and (3) $k$, the time scale constant. We set the initial conditions for the two parties as given by the data. The best fitting parameters are: $\sigma_0= 0.93$, $b= 3.73$, and $k=2.54$.
	
	\section{Discussion} 
	
	\subsection{Comparison between satisficing and maximizing models}
	
	The satisficing assumption is essential to our model's behavior. Here, we compare the results of our satisficing model with a maximizing model in the same framework (assuming two parties) and show that the predictions from the two models are fundamentally different. We consider the maximizing voters to be defined in the Downsian sense \cite{downs1957}---they maximize their utility function by voting for the ideologically closest party. With maximizing voters, the number of votes each party receives is
	\begin{equation}
	V^{(m)}_l = \int_{-\infty}^{(\mu_l+\mu_r)/2} \rho(x) dx, \; \text{and} \;V^{(m)}_r = \int_{(\mu_l+\mu_r)/2}^{\infty} \rho(x) dx \;, 
	\end{equation}
	where $(l,r) = (1, 2)$ if $\mu_1<\mu_2$, and $(l,r) = (2, 1)$ if $\mu_1>\mu_2$. If $\mu_1 = \mu_2$, both parties receive the same number of votes, 
	\begin{equation}
	V^{(m)}_1 =  V^{(m)}_2 = \frac{1}{2} \int_{-\infty}^{\infty} \rho(x) dx\;.
	\end{equation}
	We then solve for the fixed points of the system described by Eq.~\eqref{eq:dynam} with the two types of vote calculations and find their stabilities. 
	
\begin{figure}[ht]
		\centering 
		\includegraphics[width = 0.8\textwidth]{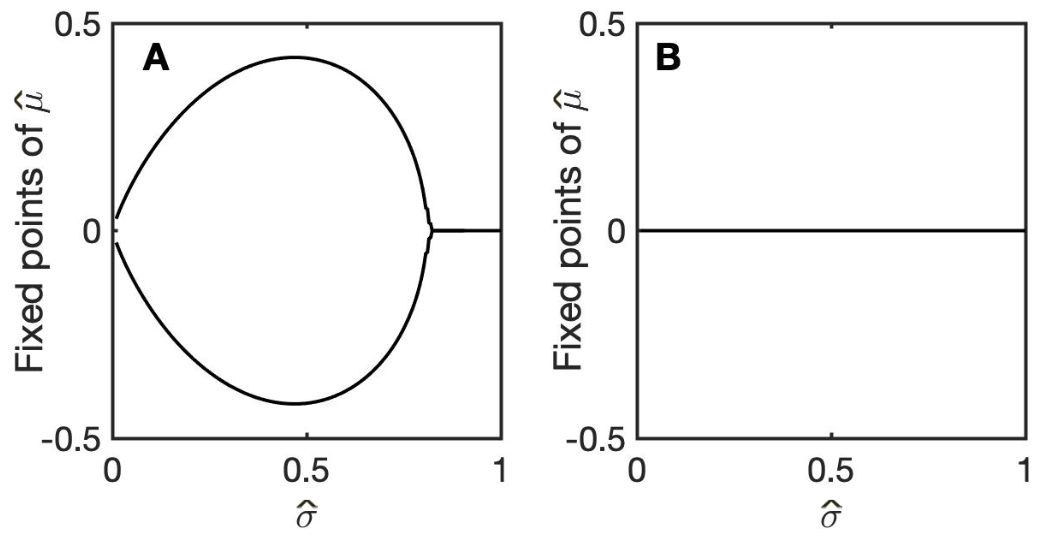}
		\caption{Stable fixed points for party positions ($\hat \mu$) as a function of the inclusiveness parameter of the two parties ($\hat \sigma$) for (A) the satisficing model and (B) the maximizing model. For each model, the positions of both parties are shown on the same vertical axis.}
		\label{figSI:max_sati_mixed}
	\end{figure}

	Figure~\ref{figSI:max_sati_mixed} shows the stable fixed points for party positions as a function of $ \sigma_1 = \sigma_2 = \sigma$, for normalized variables $\hat \sigma = \sigma/\sigma_0$ and $\hat \mu = \mu / \sigma_0$. With satisficing voters, the two parties stabilize at a finite separation for a range of parameters (as shown above). However, with maximizing voters, the position $(0,0)$ is the only stable fixed point, recovering Downs' result \cite{downs1957}, a classic finding in political science.\footnote{This result can be immediately seen for the symmetric case $\mu_1 = -\mu_2$. In that case, each party is drawn toward the center to increase votes, despite having split the number of votes equally with the opposing party.} This difference indicates that the voters' satisficing decision making is indeed a key ingredient in producing the model's results. 
	\label{secSI:mix_max_and_sat}
	
	\subsection{Implications}
	
	Our model offers new contributions to the literature. First, we present empirical evidence of and an explanation for the relationship between party homogeneity and polarization, showing that simply changing the shape of parties' satisficing functions (via changes to inclusiveness) is sufficient to lead to divergence in their positions. In particular, even though the impact of intra-party heterogeneity on a party's competitiveness has been discussed before \cite{kernell2015}, the mechanism linking increasing ideological homogeneity with diverging party positions has remained unclear. Second, the model also shows why appealing to an extreme segment of the electorate can be a winning political strategy in times of greater intra-party ideological homogeneity (i.e., decreasing party inclusiveness), which may be especially relevant for interpreting current trends in U.S.~politics. Ideological homogeneity itself may be driven by other factors, such as partisan redistricting \cite{carsonetal2010} or media echo chambers \cite{levendusky2009}, and explicitly modeling how these factors influence intra-party cohesion remains an open area of research. \add{Third, our approach offers a new quantitative framework that incorporates satisficing behavior into a \add{voting} model\del{ for voting}.} \del{Second, our model offers a quantitative framework to examine the political landscape as an outcome of satisficing voting. We incorporate this behavioral research insight into a mathematical model, which is outside of the standard framework for voting dynamics.}
	
	In addressing the call for combining interdisciplinary methods to study human social behavior \cite{gintis2004}, this article offers insight into the complex process of political elections and democratic responsiveness through a parsimonious model, \add{and suggests several directions for future work}. For example, do all individuals engage in satisficing, and if so, do they do so in the same way? When and how can minor parties gain traction? And, how might outcomes change if parties could control both their position and their level of inclusiveness? For simplicity, we left out a number of electoral variables, such as party primaries, campaign financing, and Southern realignment, but it will be important for future research to understand how these factors interact with vote satisficing. We hope our work will spur further development of quantitative frameworks to incorporate human bounded rationality into mathematical models.

	\section*{Acknowledgments}
	We thank Jeffrey Lewis and Keith Poole for in-depth discussions about DW-NOMINATE. We also gratefully acknowledge babies Miriam, Desiree, and Yara for their contributions.
	
	\subsection*{Data and code availability} 
	A copy of the DW-NOMINATE dataset used is included as a supplementary file. The computer code for numerically simulating the model is available in the following GitHub repository: \url{https://github.com/vc-yang/satisficing_election_model}.

	\section*{Author Contributions} All authors contributed to the design of the research. V.C.Y. developed the model and performed computer simulations. All authors analyzed the results and contributed to the writing of the manuscript, which was led by V.C.Y. and D.M.A.

	\section*{Funding} This work was supported by the James S. McDonnell Foundation through grant No.~220020230, the Northwestern University’s Data Science Initiative, Suzanne Hurst and Samuel Peters, and the Santa Fe Institute Omidyar Fellowship.
	

	\setcounter{figure}{0}
	\setcounter{section}{0}
	\renewcommand\thefigure{S\arabic{figure}}  
	\renewcommand\thetable{S\arabic{table}}  
	
	\newpage
	\setcounter{equation}{0} 
	\setcounter{page}{1}
	
	\begin{center}
		\Large{Supplementary Material: \\ Why Are U.S. Parties So Polarized? A ``Satisficing'' Dynamical Model}
	\end{center}
	
	\normalsize
	\begin{center}
		\author{Vicky Chuqiao Yang, Daniel M. Abrams, Georgia Kernell, and Adilson E. Motter}
	\end{center}
	
	\setcounter{equation}{0}
	\renewcommand\theequation{S\arabic{equation}}
	
	\tableofcontents

	\section{Additional discussion of model assumptions}
	
	\add{We discuss the empirical basis for the assumed ideology distribution of the voting population and key properties of a higher-dimensional extension of the model.}
	\label{secSI:assumptions}
	\subsection{Population ideology distribution} 
	
	In the main text, we assumed the public's ideology distribution to be unimodal and approximated it as Gaussian. Here we use empirical data to justify this assumption. 
	
\begin{figure}[!htb]
		\centering 
		\includegraphics[width = 0.7\textwidth]{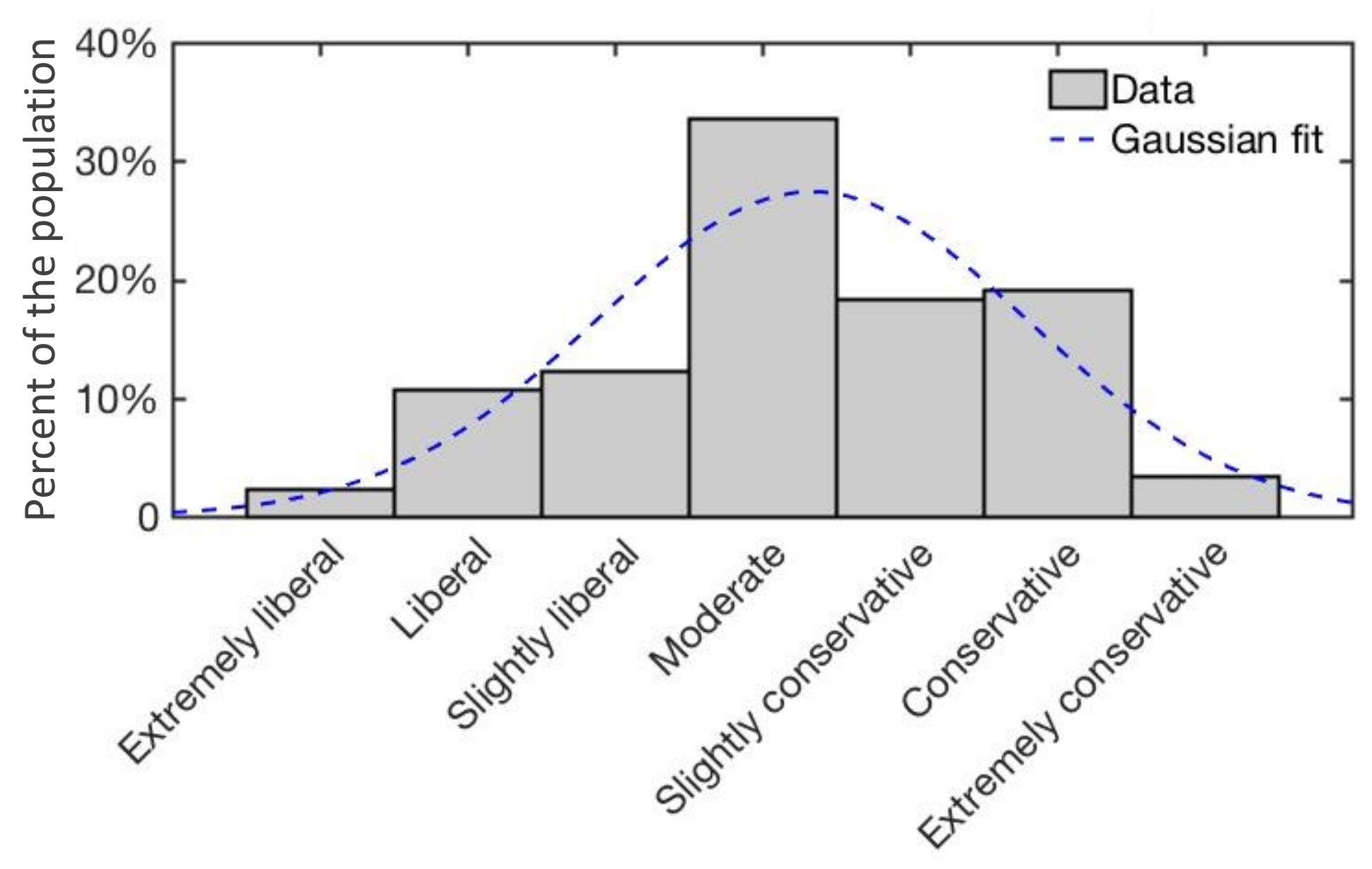}
		\caption{Distribution of self-reported ideology in \add{the} ANES data (1972--2012 aggregated) compared with a Gaussian fit. The $\text{R}^2$ of the Gaussian fit is 0.74.}
		\label{figSI:ANES_data}
	\end{figure}

	The American National Elections Studies (ANES) conducted yearly surveys from 1972 to 2012 asking people's self-identification on a 7-point scale between extremely liberal and extremely conservative. The self identification shows a unimodal distribution centered at ``moderate.'' The data were downloaded in November 2015 from \url{http://www.electionstudies.org/studypages/anes_timeseries_cdf/anes_timeseries_cdf.htm}. \\The histogram for the U.S. public self-identified ideology is shown in Fig.~\ref{figSI:ANES_data}. Here\add{,} the distribution remains peaked at the moderate position for all years, despite fluctuations, and the Gaussian distribution is a good approximation of this distribution. 
	
	\subsection{\add{Dimensionality of the} ideology space} \label{secSI:2D}
\begin{figure}[ht]
		\centering 
		\includegraphics[width = 0.6\textwidth]{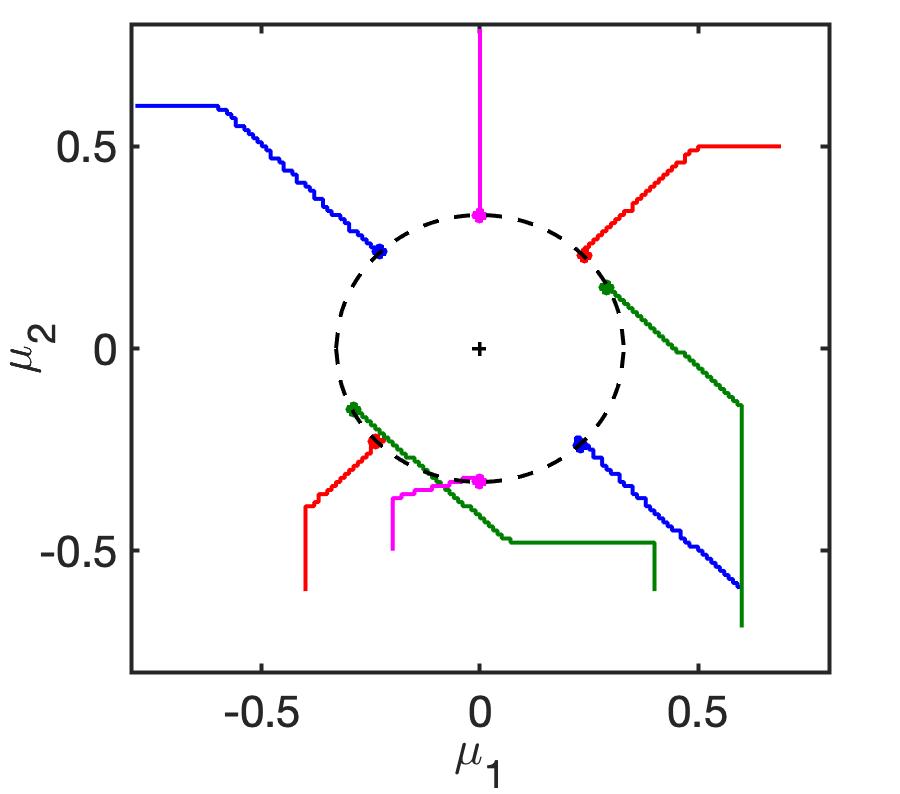}		
		\caption{Example trajectories of the two-party system in a two-dimensional ideology space, where \add{ideological positions are} denoted by the coordinate\add{s} \add{$(\mu_1, \mu_2)$}. Solid circles denote the steady-state positions of the parties and \add{the colored curves} denote transient trajectories from various initial conditions. \add{Curves of the same color indicate the} two parties' trajectories from one set of initial conditions. In \add{these simulations}, the parameters are \add{set to} $\sigma_0 = 1$ \add{and} $\sigma_{1, 2} = 0.2$. For simplicity, the simulation is restricted to parties moving parallel to the horizontal or vertical axis at each time step. \add{The dashed circle is included to facilitate visualization of the steady-states positions.}}
		\label{figSI:2d_traj}
	\end{figure}

	In the main text\add{,} we are motivated by the reality that political idealogy in the U.S.~is largely one-dimensional. The model can be generalized to higher dimensions by replacing \add{the} scalar variables $x$, $\mu_i$, and $\sigma_i$ with vectors. \add{Figure} \ref{figSI:2d_traj} shows some example trajectories for the two-party case in a two-dimensional \add{ideology space}. In these simulations, for ease of numerical computation, we used discrete time steps and restricted party movements in each period to either vertical or horizontal directions, but the steady states \add{reached are equivalent to those found by} integrating the differential equations. The behavior of the system is analogous to that of the one-dimensional problem.  For large \add{values of the} parameters $\sigma_i$, both parties converge to the center. \add{For small-to-moderate $\sigma_i$, however,} the parties stabilize at a finite separation on opposite sides of a circle (Fig.~\ref{figSI:2d_traj}). 
	
	\section{Data for \add{the} U.S.~Democratic and Republican party positions}
	\label{secSI:nominate_data}
	We use congressional voting records compiled with the \add{Dynamic, Weighted, Nominal Three-Step Estimation (DW-NOMINATE)} algorithm to empirically measure the party positions, $\mu$, and the inclusiveness parameter, $\sigma$, for the Democratic and Republican parties. DW-NOMINATE is a multi-dimensional scaling method that first calculates a pairwise distance for every two members of Congress based on similarity in their roll call vote records. It then projects the resulting high-dimensional network of \del{congressional members}\add{legislators} to a \add{low-dimensional space} while preserving the \add{pairwise}\del{pair-wise} distance relation as much as possible. The \del{representatives'}\add{legislators'} relative positions in this \add{low-dimensional space} are referred to as their ideology scores. This \add{procedure} was conducted for the House and Senate separately. The ideology scores of the two chambers \add{were then combined} into one dataset, \add{which is the dataset} used in our \add{analyses}. 
	
	We used the version of the House and Senate combined dataset that was last updated in 2015. The data was downloaded from \url{https://legacy.voteview.com/dwnomin.htm} in May 2016. The website has since been updated. The new link (as of September 2018) for data download is \url{https://voteview.com/data}. A copy of the dataset that the authors downloaded in May 2016 and used in this analysis can be found \add{as a} supplementary data file. 
	
	According to Poole and Rosenthal \cite{poole2011}, despite \add{the underlying} complexity, \add{the} roll call votes in the House and the Senate can be organized and explained by no more than two dimensions throughout American history. The first dimension, also called the dominant dimension, is commonly thought of as tapping into the left-right (or liberal--conservative) spectrum on economic matters. The second dimension picks up attitudes on cross-cutting and salient issues (e.g., slavery, civil rights, and social/lifestyle issues). 
	
	We use the DW-NOMINATE first dimension score to measure a \del{congressmember's} \add{legislator's} ideological position. To estimate each party's position, $\mu_i$, we use the mean position of a party's \del{congressmembers} \add{legislators}. We also estimate each party's inclusiveness parameter, $\sigma_i$, as proportional to the standard deviation of \del{congressmembers'}\add{the legislators'} positions. The second dimension accounts for at most an additional 6\%, and at the lowest point an additional 1\%, of explanatory power in legislators' votes \cite{poole2011}.

	\section{Robustness checks} \label{secSI:robustness}
	In this section, we provide evidence that the core conclusions of our paper do not change when using alternative datasets and slightly different model assumptions. 
	
	\subsection{Validation for the satisficing function from independent data}
	For a robustness check of our findings, we \add{employed} an alternative estimate of the satisficing function parameter $\sigma$ using a \add{dataset that is} independent of the DW-NOMINATE scores shown in the main text. 
	
	The data used is the American National Elections Studies (ANES) 1948-2012 time series dataset, as described in section 1 above. We used three variables from the survey: \add{the} self-reported liberal-conservative scale (VCF0803), \add{the} feeling thermometer for liberals (VCF0211), and \add{the} feeling thermometer for conservatives (VCF0212). Data on these variables are available for 18 years between 1972 \add{and} 2012, for 55,674 individuals in total. 
	
	The feeling thermometer questions ask participants to report their feeling towards a group on a 0-100 scale. Participants are told that 50--100 means they feel favorably towards the group, 0--50 means they do not feel favorably towards the group, and 50 means they do not feel particularly warm or cold. 
	
\begin{figure}[ht]
		\centering
		\includegraphics[width=0.8\textwidth]{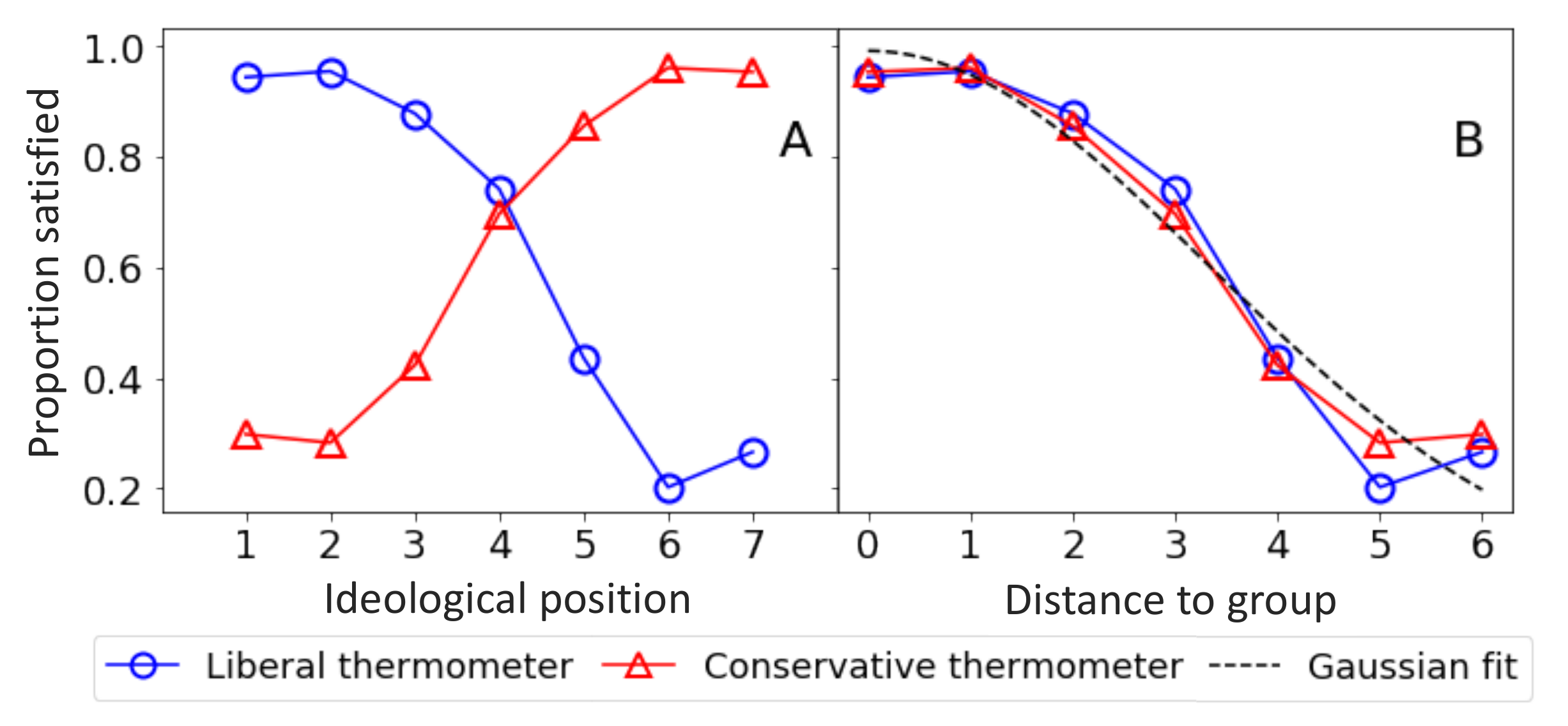}
		\caption{ANES thermometer data from 2012. We assume \add{that} thermometer $\geq 50$ means satisfied. (A) Proportion of the population satisfied with liberals and conservatives by \add{their} ideological position, \add{where 1 means extremely liberal and 7 means extremely conservative} (B) Proportion satisfied as a function of the distance to the target groups (liberals or conservatives), and Gaussian fit. The two curves collapse and can be approximated by a Gaussian function. The best fitting Gaussian has \add{a} standard deviation \add{of} $3.3$. }
		\label{figsi:ANES_sigma_2012}
	\end{figure}

	We assume \add{that} thermometer $\geq 50$ means satisfied with the liberal/conservative position. The self-reported liberal\add{-}conservative scale is 1 to 7, where 1 represents most liberal, and 7 most conservative. \add{Fig.}~\ref{figSI:ANES_data} shows the aggregated distribution of the data from 1972 to 2012. Given the definition of the \add{7}-point scale, we interpret the target group (liberal and conservative) as positioned at 1 and 7 of this scale, respectively. Then\add{,} we calculate the proportion of the population satisfied with the target group as a function of \add{their} distance to the group. We fit a Gaussian to this curve and get the parameter $\sigma$ from the fit, which corresponds to the width of the satisficing function. Fig.~\ref{figsi:ANES_sigma_2012} shows\add{, as an example,} the behavior of the data in 2012. The two satisficing curves for conservatives and liberals collapse, and \add{are} well approximated by a Gaussian function. 
	
	\begin{figure}[ht]
		\centering
		\includegraphics[width=0.8\textwidth]{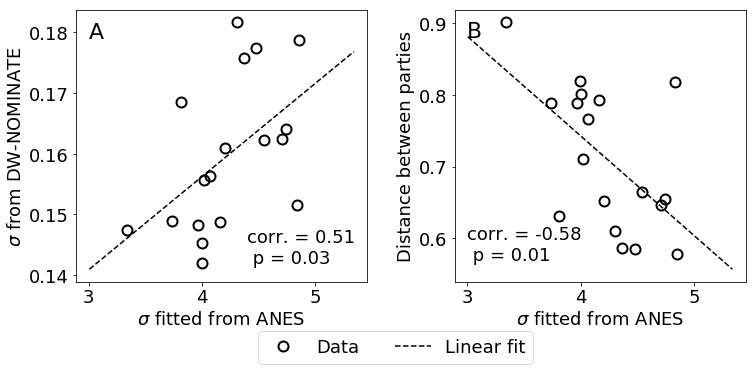}
		\caption{Comparing $\sigma$ fitted from \add{the} ANES data and metrics computed from \add{the} DW-NOMINATE data. Left: the $\sigma$ parameters inferred from the two sources are positively correlated, giving us some confidence in using the DW-NOMINATE standard deviation to approximate the ``tolerance'' of voters. Right: \add{the} parameter $\sigma$ fitted from \add{the} ANES data is negatively correlated with the distance between parties \add{(i.e., party polarization)}, which \add{supports the} observation made in the main text using \add{$\sigma$ derived from the DW-NOMINATE data}.}
		\label{figsi:ANES_sigma_corr_nominate}
	\end{figure}
	
	We repeat the analysis for all 18 years of data available. \add{Fig}.~\ref{figsi:ANES_sigma_corr_nominate} plots the best fitting $\sigma$ value to the ANES data of each year (horizontal axis) against the $\sigma$ value estimated from the DW-NOMINATE data of the Congress that includes the same year. The Pearson correlation is $0.51$ \add{(}$p = 0.03$\add{)}. The $\sigma$ value estimated from ANES correlates with distance between parties \add{(i.e., party polarization)} in \add{the} DW-NOMINATE data (Congress data) with Pearson correlation $-0.58$ \add{(}$p = 0.01$\add{)}.
	
	This analysis serves as a robustness check for the negative relationship between party separation (i.e., political polarization) and the narrowing of the satisficing function (i.e., increasing intra-party ideological homogeneity or \add{reducing} inclusiveness). 
	
	\subsection{Polarization in alternative metrics of party position}\label{subsec:alt_metric}
	The DW-\\NOMINATE score has been criticized at times for its computational complexity and for the difficulty of interpreting its axes.  Here, we present \add{a simpler and easier-to-understand metric} and show that a negative relationship between polarization and inclusiveness is still observed. 
	
	We use data from the U.S.~House of Representatives' roll call records \add{for the period} 1941--2015. \add{The data} were downloaded from \url{https://www.govtrack.us/data/} in March 2016. 
	
	We calculate two scores for each representative. For each representative in each Congress, we calculate a Republican alignment score ($S_R$). \add{This score measures} the proportion of time\add{s} she or he voted in agreement with the Republican majority position. For example, if 60\% of Republican representatives supported a bill, then a representative voting in favor of that bill would be counted as voting in agreement with \add{the} Republican majority. More explicitly, 
	\begin{equation}
	S_R = \frac{N_R}{N},
	\end{equation}
	where $S_R$ is the \add{Republican alignment} score, $N_R$ is the number of times voting with the Republican majority, and $N$ is the total number of votes. Similarly, we also calculate a Democratic alignment score for each representative,  
	\begin{equation}
	S_D = \frac{N_D}{N},
	\end{equation}
	where \add{$N_D$} is the number of times a representative votes with the \add{Democratic} majority. 
	
\begin{figure}[thb]
		\centering 
		\includegraphics[width = 0.9\textwidth]{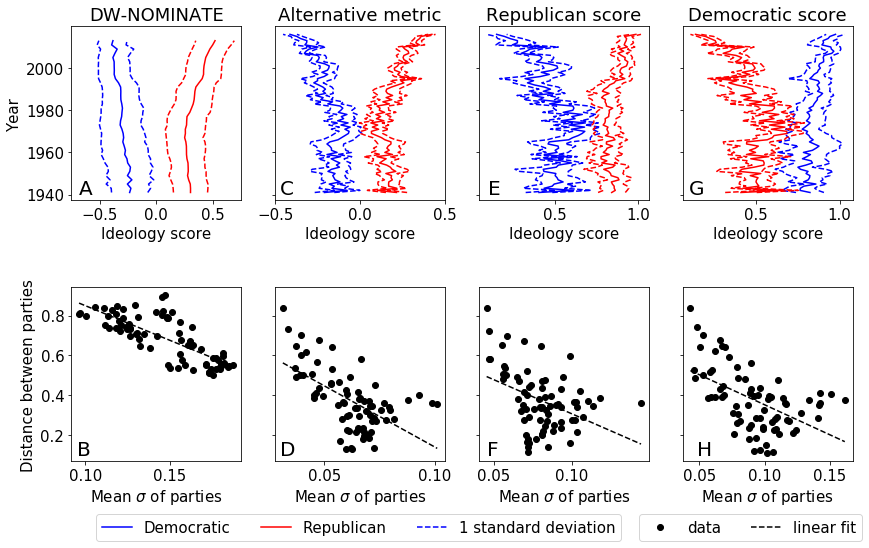}
		\caption{Party polarization trends from alternative metrics. Top row\add{:} time history of party positions and standard deviations measured by the various metrics. Bottom row\add{:} distance between parties \add{(i.e., party polarization)} vs.~party inclusiveness based on various metrics. (A, B) DW-NOMINATE data (same as used in Fig.~4 of the main text). (C, D) Alternative metric $S$. (E, F) Republican alignment score $S_R$. (G, H) Democratic \add{alignment} score $S_D$. }
		\label{figSI:alt_metric_res}
	\end{figure}

	Finally, we combine the two \add{scores} into one alternative metric\add{:}
	\begin{equation}
	S =  \frac{1}{2} (S_R- S_D).
	\end{equation}
	Results of this analysis are shown in Fig.~\ref{figSI:alt_metric_res}. The polarization trend is observable even \add{using} this simple metric (shown in the \add{second} column of Fig.~\ref{figSI:alt_metric_res}). The negative relationship \add{between the interparty distance (i.e., party polarization)} and party inclusiveness is still present. The Pearson correlation in the DW-NOMINATE data is $-0.77$ \add{(}$p = 2\times 10^{-16}$\add{)}, and the Pearson correlation in the alternative metric is $-0.59$ \add{(}$p = 2 \times 10^{-8}$\add{)}. Note that the Republican and Democratic scores (shown in the \add{third} and \add{fourth} columns of Fig.~\ref{figSI:alt_metric_res}) exhibit asymmetry due to a number of unanimous votes. 
	
	\section{Alternative model that maximizes vote share}
	\label{secSI:vote_share_model}
	
	In Eq.~(3.3) of the main text, we assume that each party acts to maximize its number of votes. But in many cases parties may seek instead to maximize their vote 
	\textit{share}. We analyze this alternative model and find that \add{similar qualitative conclusions hold; specifically, lower inclusiveness is related to greater party polarization.} \add{In the main text we chose to present the vote-maximizing model because it requires fewer assumptions and parameters. }
	
	In \add{the} alternative model for elections \add{among parties $1$ through $n$}, Equation (3.3) is modified to
	\begin{equation}\label{eq:dynam_max_share}
	\d {\mu_i} t  = k \pd{}{\mu_i} \left(\frac{V_i}{\sum_{j = 1}^n V_j}\right) \;,
	\end{equation}
	
	where $n$ is the number of parties. Similar expressions for $p$ can be derived for multi-party systems. For example, in a three-party system,   
	\begin{equation} \label{eqSI:p3_1}
	p^{}_{1} = \underbrace{s_1 (1 - s_2)(1 - s_3)}_\text{term 1} +  \frac{1}{2} \underbrace{\left(s_1 s_2(1 - s_3) +  s_1 s_3 (1 - s_2)\right)}_\text{term 2}  + \frac{1}{3} \underbrace{s_1 s_2 s_3}_\text{term 3}\;, 
	\end{equation}
	where the functions $s_1$, $s_2$, and $s_3$ are the same satisficing functions as described in the main text. Term 1 gives the expected proportion of voters satisfied with party $1$ only. Term 2 gives the proportion satisfied with party $1$ and exactly one other party. Term 3 gives the proportion satisfied with all three parties.  
	
\begin{figure}[thb]
		\centering 
		\includegraphics[width = 0.6\textwidth]{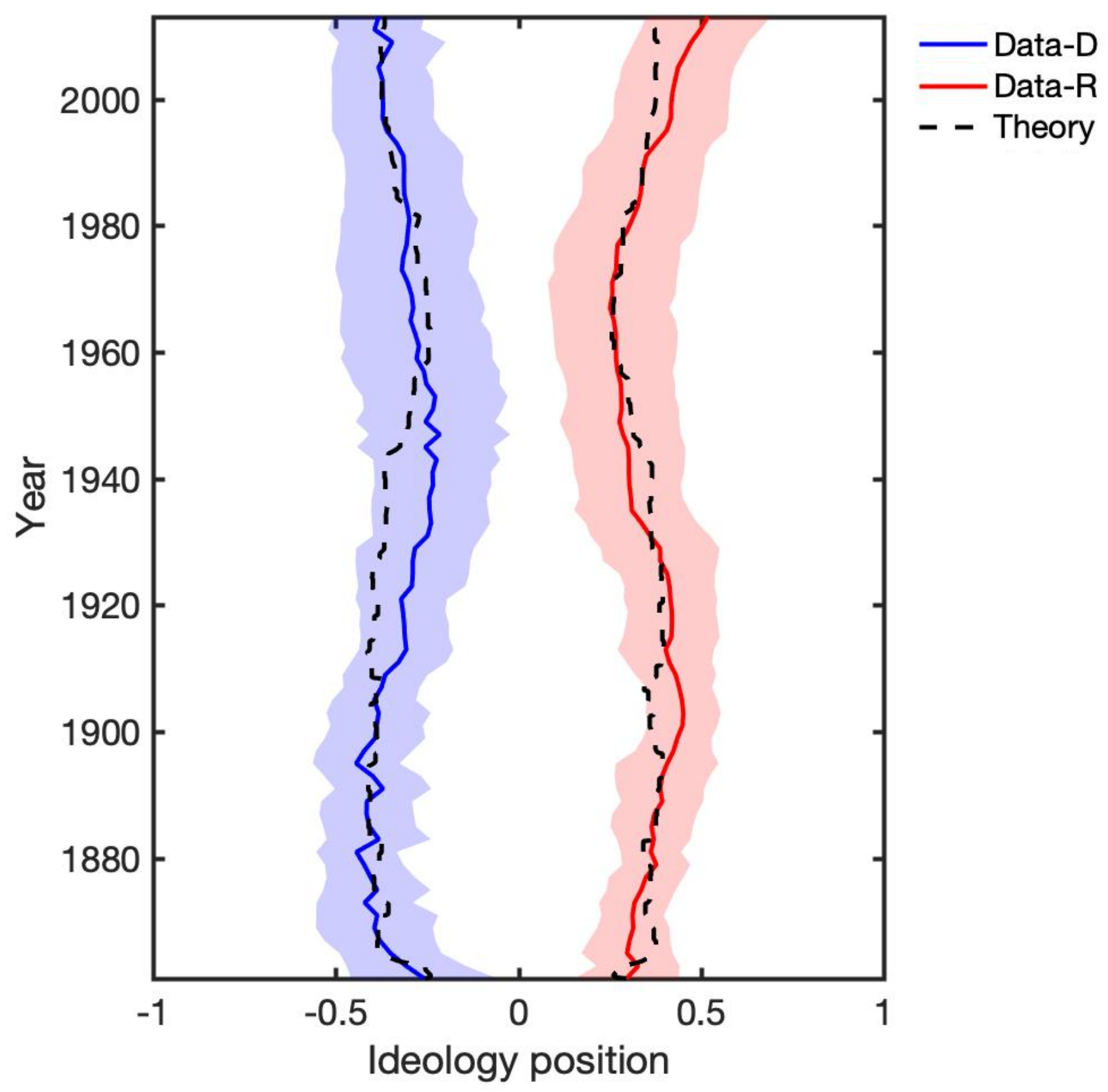}
		\caption{\add{The comparison of party position over time predicted by the alternative model for maximizing vote share with empirical data. In the alternative model, we assume two major parties (shown), and two minor parties on either side of the major parties (not shown).}}
		\label{figSI:fig4b_share_model}
	\end{figure}

	In the case where exactly two parties compete, Eq.~\eqref{eq:dynam_max_share} predicts that both parties will converge to the median voter's position, recovering a classic result in political science \cite{downs1957S}. However, this strategy is sensitive to the presence of even very small third parties. In the multi-party version of Eq.~(3.3), with \add{a minor third party} (characterized by \add{a small $\sigma_3$} parameter), the outcome changes: the two major parties diverge from the median and stabilize at a finite \add{distance from each other}. Intuitively, because voters who are unsatisfied with either party choose to abstain, in a strictly two-party system, the median voter's position does not maximize the number of votes for either party, although it does maximize the vote share. It is only with the threat of additional parties capturing lost votes that convergence to the median ceases to be optimal. This prediction is in agreement with previous research that finds that third-party candidates can shape election outcomes \cite{lee2012}. We also perform analyses similar to those shown in Fig.~4 using a four-party version of Eq.~(3.3), consisting of two major parties and two minor parties, and find qualitatively similar results \add{(an example is shown in Fig.~\ref{figSI:fig4b_share_model}).}
	
\end{document}